\documentstyle[12pt]{article}
\def \beq{\begin{equation}}
\def \eeq{\end{equation}}
\def \beqn{\begin{eqnarray}}
\def \eeqn{\end{eqnarray}}
\def \r{{\bf r}}
\def \T{{\bf T}}
\def \semidir{\bowtie}

\begin{document}
\begin{titlepage}
\begin{flushright}\vbox{\begin{tabular}{c}
           TIFR/TH/99-07\\
           March, 1999\\
           hep-lat/9903019\\
\end{tabular}}\end{flushright}
\begin{center}
   {\large \bf
      Representations of Fermionic Correlators\\
      at Finite Temperatures}
\end{center}
\bigskip
\begin{center}
   {Sourendu Gupta\footnote{E-mail: sgupta@tifr.res.in}\\
    Department of Theoretical Physics,\\
    Tata Institute of Fundamental Research,\\
    Homi Bhabha Road, Mumbai 400005, India.}
\end{center}
\bigskip
\begin{abstract}
The symmetry group of the staggered Fermion transfer matrix in a spatial
direction is constructed at finite temperature. Hadron-like operators
carrying irreducible representations of this group are written down from
the breaking of the zero temperature group. Analysis of the correlators
in a free fermion theory suggests new measurements which can test current
interpretations.
\end{abstract}
\end{titlepage}

\section{\label{sc.intro}Introduction}

Lattice simulations of field theories in equilibrium at finite temperature
($T$) use a discretisation of the Euclidean formulation for partition
functions---
\begin{equation}
   Z(\beta)\;=\;\int{\cal D}\phi\exp\left[-\int_0^\beta dt\int d^3x
      {\cal L}(\phi)\right],
\label{int.part}\end{equation}
where $\phi$ is a generic field, $\cal L$ the Lagrangian density, and
the Euclidean ``time'' runs from 0 to $\beta=1/T$. The path integral
is over Bosonic (Fermionic) field configurations which are periodic
(anti-periodic) in Euclidean time. Due to the lack of symmetry between
the space and Euclidean time directions in eq.\ (\ref{int.part}), this
problem has only a subgroup of the full 4-dimensional rotational
symmetry of the $T=0$ Euclidean theory. In this paper we focus on the
lattice discretised problem, where all continuum symmetries break to a
discrete subgroup.

It is possible to write the partition function of eq.\ (\ref{int.part})
as the trace of the transfer matrix in one of the spatial directions.
The symmetry groups we examine leave such a transfer matrix invariant.
The eigenvectors of the matrix carry irreducible representations (irreps)
of these symmetry groups.

Thermodynamics depends only on the leading eigenvalue, which always belongs
to a scalar representation of the symmetry group. Hence the group theory is
not crucial for the study of properties such as the phase structure,
transition temperature, $T_c$, and other thermodynamic quantities. In fact,
extensive measurements have been made of $T_c$ for pure gauge theories, and
those with massless Fermions \cite{chiral}, and our group theoretical analysis
adds very little to this.

However, the symmetry properties are crucial to the study of screening
correlation functions and the determination of screening masses. These can be
written in terms of the ratio of the largest and an appropriate other
eigenvalue of the transfer matrix. The significance of the equality (or
otherwise) of two screening masses will depend on whether or not the
correlation function lies in the same irrep of the symmetry group of the
transfer matrix. In the gauge sector of the theory this analysis has been
carried out and applied to the study of screening masses \cite{old}, to
demonstrate dimensional reduction in a fully non-perturbative manner.

Screening masses obtained from correlation functions built out of staggered
Fermion field operators have also been extensively studied in the past
\cite{screen,fss,myold}. Screening masses in the high temperature phase of
QCD seem to approach those expected from free field theory as $T\to2T_c$
\cite{fss,myold,milc}. Some other measurements which seem to indicate that
the picture may be more complicated \cite{milc}, also turn out to be
explained in terms of weakly interacting quarks \cite{suny}.
All these studies have relied entirely on the $T=0$ analysis of the lattice
symmetry group of staggered Fermions.

In this paper we present the
first analysis of the symmetries of the corresponding finite temperature
problem. We find that all the screening masses measured till now see only
one of the representations of the symmetry group. Many
other masses can be studied, and are likely to yield further information
about the theory. The free field theory of these other representations
is worked out.

One observation arising from the application of these group theoretical
results to previous simulations is worth mentioning in the introduction.
Since the $T=0$ scalar and pseudo-scalar mesons, and the symmetric linear
combinations of the three components of the vector and pseudo-vector lie
in the same irreducible representation (irrep) of the point group of a
$T>0$ spatial slice through the lattice, they must have degenerate masses
in a free fermion theory. When interactions switch on, the relevant symmetry
becomes that of an enveloping group, and the four degenerate masses split
into two pairs of degenerate masses. Observation of such a splitting for
$T\le2T_c$ \cite{screen,fss,myold,milc} must then be interpreted as evidence
for interactions \cite{myold}. Nothing further can be said purely from the
study of these correlators. Whether the spectrum of screening masses comes
from a weakly interacting effective theory, or whether it is very similar
to the spectrum at zero temperature, are questions which can only be
answered by measuring the masses in the other representations which we
write down explicitly.

In Section \ref{sc.rev} we present a brief review of the symmetries of
staggered Fermions at $T=0$. This serves to set up the notation, and
indicates what changes to expect at finite temperature.
Section \ref{sc.reps} contains our main results on the characterisation of
the group of symmetries of the spatial transfer matrix at $T>0$ and its
irreducible representations (irreps). Free field theory results for the
screening masses is discussed in Section \ref{sc.fft},
where presently available data is also discussed.
Two appendices contain the technical details of induced representations
and character tables for the irreps of mesons.

\section{\label{sc.rev}Symmetries of Staggered Fermions}

In this section we review the breaking of continuum spin-flavour symmetries
for lattice staggered Fermions \cite{ksf} at zero temperature, and identify
how this pattern changes at finite temperature. The continuum symmetry for
four flavours of Fermions is $SU_r(2)\otimes U_f(4)$, where the first factor
is the rotational symmetry, and the second is the flavour symmetry. We follow
the notational conventions of \cite{golt1,golt2}.

At $T=0$ we are interested in the symmetries of Fermion operators which have
zero momentum in the directions orthogonal to the Euclidean time $t$---
\begin{equation}
   \chi_A\;=\;\sum_{{\bf m}}T_z^{-m_3}T_y^{-m_2}T_x^{-m_1}
      \chi(r)T_x^{m_1}T_y^{m_2}T_z^{m_3} \;=\;
    \sum_{{\bf m}}\chi({\bf x}+2{\bf m}a).
\label{rev.zerom}\end{equation}
Here the index $A$ denotes the corners of the hypercube on which the
appropriate component of the quark field resides, $a$ is the lattice
spacing, $T_i$ are the generators of translations in the $i$-th direction,
and we have assumed that there are periodic boundary conditions in all
directions on the slice. In writing eq.\ (\ref{rev.zerom}), we have
chosen to study correlation functions of operators separated in the time
direction. Due to the 4-dimensional discrete rotational symmetry of the $T=0$
theory, we could have chosen to study propagation in any other direction
with the same result.

\begin{table}[htbp]\begin{center}\begin{tabular}{|c|c|}
\hline
Operation & Action \\
\hline
$R_{\kappa\lambda}$ & $\chi(x)\to {\cal R}(R_{\kappa\lambda}^{-1}x)\,\chi(R_{\kappa\lambda}^{-1}x)$ \\
 &    ${\cal R}(x)\;=\;{1\over2}\biggl[1\pm\eta_\kappa(x)\eta_\lambda(x)
          \mp\zeta_\kappa(x)\zeta_\lambda(x)+$ \\
 &    $\qquad\ \ \ +\eta_\kappa(x)\eta_\lambda(x)
        \zeta_\kappa(x)\zeta_\lambda(x)\biggr]$ \\
$I$ & $\chi(x)\to\eta_4(x)\chi(I x)$ \\
$S_\mu$ & $\chi(x)\to\zeta_\mu(x)\chi(x+a_\mu)$ \\
$C$ & $\chi(x)\to\epsilon(x)\chi(x)$ \\
$U_B(1)$ & $\chi(x)\to {\rm e}^{i\Theta_B}\chi(x)$ \\
\hline
\end{tabular}\end{center}
\caption[dummy]{Symmetry operations on staggered Fermions. The upper
   (lower) signs in ${\cal R}$ are used when $\kappa>\lambda$ ($\kappa<\lambda$).
   Here $\epsilon(x)=(-1)^{x_1+x_2+x_3+x_4}$, $\eta_i(x)=\prod_{k<i}
   (-1)^{x_k}$ and $\zeta_i(x)=\prod_{k>i}(-1)^{x_k}$.}
\label{tb.opers}\end{table}

The symmetry elements of the theory are listed in Table \ref{tb.opers}.
For staggered Fermions, the shifts by one lattice spacing, $S_\mu$, are
mixed flavour and translation operations. Pure translations are $T_\mu=
S_\mu^2$. We have chosen the transfer matrix $\T$ to be $T_t$. Nothing
would have changed, at $T=0$, if we had instead chosen $\T$ to be $T_z$.

Discrete flavour operations, $\Xi_\mu=S_\mu T_\mu^{-1/2}$, are vectors
under rotations, generated by $R_{\kappa\lambda}$ and transform as
\begin{equation}
  R_{ij}^{-1}\Xi_k R_{ij}\;=\; \delta_{ik}\Xi_j + \delta_{jk}\Xi_i
          +|\epsilon_{ijk}|\Xi_k.
\label{rev.flav}\end{equation}
Here and elsewhere, Greek indices run from 1 to 4; Latin indices over the
three directions summed in eq.\ (\ref{rev.zerom}) or its analogue.
A subgroup $U_B(1)$ of the continuum flavour group remains unbroken on the
lattice; this charge corresponds to the Fermion number $q$. The
representations of $\Xi_\mu$ in an irrep with Fermion number $q$,
${\cal D}_q(\Xi_\mu)$, obey the relation
\begin{equation}
   {\cal D}_q(\Xi_\mu) {\cal D}_q(\Xi_\nu)\;=\;{\rm e}^{i\pi q}
      {\cal D}_q(\Xi_\nu) {\cal D}_q(\Xi_\mu).
\end{equation}
Inversion, $I$, commutes with $\Xi_4$, and anti-commutes (commutes) with
the other $\Xi_k$ in representations with odd (even) values of $q$. Parity
is defined by $P=\Xi_4I$. The remaining discrete symmetry is that of
charge-conjugation, $C$.

The symmetries of $\T$ are the rest-frame group
\begin{equation}
   RF\left(\Xi_\mu,R_{kl},I,C\right)\otimes U_B(1).
\label{rev.rf}\end{equation}
We have used the notation $G(X)$ to mean the group $G$ generated by the
operation(s) $X$. A subgroup of $RF$ is the group of isometries of the
lattice, called the geometric rest frame group, $GRF\left(\Xi_\mu,R_{kl},
I\right)$. In turn, $GRF$ contains the time slice group, which is the
point group of the lattice---
\begin{equation}
   TS\left(R_{kl},I\right)=O_h\left(R_{kl},I\right)=
      O\left(R_{kl}\right)\otimes Z_2(I).
\label{rev.ts}\end{equation}
This chain of groups builds up to the continuum symmetry group---
\begin{equation}
   TS\subset GRF\subset RF\subset SU_d(2)\otimes U_B(1),
\label{rev.chain}\end{equation}
where $SU_d(2)$ is the diagonal subgroup of the direct product
$SU_r(2)\otimes SU_f(2)$ of rotations and flavour. The breaking of $SU_f(4)$
to $SU_f(2)$ is specified by requiring that the fundamental of $SU_f(4)$
break into the irrep $({1\over2},{1\over2})$ of $SU_f(2)$.

All correlation functions block diagonalise into irreps of $TS$. This
group, $O_h$, is the group of symmetries of a cube. It has 48 elements
in 10 conjugacy classes \cite{hammer}. It has four one-dimensional irreps
$A_1^\pm$ and $A_2^\pm$, two two-dimensional irreps $E^\pm$ and four
three-dimensional irreps $F_1^\pm$ and $F_2^\pm$. The physical
interpretation of each mass is obtained by tracing the descent of the
irrep of $TS$ through the whole chain in eq.\ (\ref{rev.chain}) from the
irreps of the continuum symmetry,
$SO(4)\otimes SU(4)$. This is done in \cite{golt1,golt2}. See also
\cite{edwin} for some details of the treatment of correlation functions.

For the study of equilibrium finite temperature, $T>0$, physics we are
interested in screening masses and screening correlation functions,
{\sl i.e.\/}, in the eigenvalues of the transfer matrix in spatial
directions. Two distinctions from the $T=0$ theory should be borne in
mind.

The first is that there are anti-periodic boundary conditions
in the Euclidean time direction on Fermions. As a result the lowest Fourier
component has a non-vanishing momentum in this direction---
\begin{equation}
   T_t^{-N_t/2}\chi(x)T_t^{N_t/2}={\rm e}^{i\pi}\chi(x),
       \qquad
   T_t^{-N_t/2}\bar\chi(x)T_t^{N_t/2}={\rm e}^{-i\pi}\bar\chi(x),
\label{rev.matsub}\end{equation}
where $N_t$ is the number of lattice points in the time direction. This is
a trivial change. For Fermion bilinear operators it makes no difference.
Operators with an odd number of Fermion fields are treated slightly
differently. For example, the projection on the lowest momentum state of a
Fermion field is not written as in eq.\ (\ref{rev.zerom}), but as
\begin{equation}
   \chi_A\;=\;\sum_{{\bf m}}{\rm e}^{2i\pi m_t/N_t}\chi({\bf x}+2{\bf m}a),
\label{rev.field}\end{equation}
where ${\bf m}$ runs over the coordinates in a spatial slice, {\sl i.e.\/},
over two spatial directions and the temporal direction \cite{fss}. The phase
factor in the sum is just the statement that the lowest Matsubara frequency
for Fermions is $\pi T$.

In this paper we shall concern ourselves with the second, and more
important, difference--- the isometries of a slice of the lattice. Since we
are interested in screening masses, we consider slices through the
lattice orthogonal to one of the spatial directions, say the $z$-direction
(as in eq.\ \ref{rev.field} above).
Then the isometries of the $z$-slice generate
\begin{equation}
  \underline{TS}=D_4^h=D_4\left(R_{xy},R^2_{xt}\right)\otimes Z_2(I).
\label{rec.tts}\end{equation}
The identification of this group is easy, because it differs from $O_h$
(eq.\ \ref{rev.ts}) by the fact that rotations of $\pi/2$ in the $xt$ and
$yt$ planes is not allowed. The spectrum of the screening masses requires
a classification by the irreps of $D_4^h$. The continuum symmetry will be
built up by the group chain
\begin{equation}
   \underline{TS}\subset\underline{GRF}
    \subset\underline{RF}\subset{\cal C}\otimes U_B(1)
     \subset SU_d(2)\otimes U_B(1),
\label{rev.tchain}\end{equation}
where ${\cal C}=O(2)\otimes Z_2(I)$ is the invariance group of a cylinder.
To generate each of the lattice groups in the chain, we use the construction
at $T=0$, only leaving out odd powers of the rotations $R_{kt}$.

$D_4^h$ has 16 elements in six conjugacy classes \cite{hammer}.
There are eight one-dimensional irreps labelled $A_1^\pm$, $A_2^\pm$, $B_1^\pm$
and $B_2^\pm$, and two two-dimensional irreps $E^\pm$. The reductions of
the irreps of $O_h$ to $D_4^h$ is as---
\beqn
\nonumber
   &A_1^P\;\to\;A_1^P,\qquad&\qquad A_2^P\;\to\;B_1^P,\\
\nonumber
   &F_1^P\;\to\;A_2^P\oplus E^P,&\qquad F_2^P\;\to\;B_2^P\oplus E^P,\\
   &E^P\;\to\;A_1^P\oplus B_1^P.&
\label{rev.oh2d4h}\eeqn
More details can be found in \cite{old}.

In the rest of this paper we shall give these decompositions of 
meson and hadron operators using the language of the $T=0$ theory. This
calls for some care in the interpretation of results--- although we shall
talk of charge conjugation, $C$, and parity, $P$, and the operators will
have the same structure and algebra as in the $T=0$ theory, they may
represent quite different physical quantities \cite{yaffe}.

\section{\label{sc.reps}The Symmetry Group at $T>0$.}

In this section the symmetry groups are written down. The representation
theory of these groups in the meson (quark-antiquark) sector is examined
in detail. The symmetries of the quark fields are also examined briefly,
and the representation theory in the baryon sector is dealt with in less
detail.

\subsection{\label{sc.mesons}Meson Operators}

\begin{table}[htbp]\begin{center}
  \begin{tabular}{|c|c|c|c|c|}  \hline
  $G$ & $O$ & $\check G$ & $D_4$ & Meson \\
  \hline
  ${\bf 1}$      & $A_1$     & ${\bf 1}_0$            & $A_1$        & Yes \\
  ${\bf 1}'$     & $A_2$     & ${\bf 1}_1$            & $B_1$        &\\
  ${\bf 2}$      & $E$       & ${\bf 1}_0+{\bf 1}_1$  & $A_1+B_1$    &\\
  ${\bf 3}$      & $F_1$     & ${\bf 2}_2+{\bf 1}_6$  & $A_2+E$      & Yes \\
  ${\bf 3}'$     & $F_2$     & ${\bf 2}_2+{\bf 1}_7$  & $B_2+E$      &\\
  ${\bf 3}''$    & $F_1$     & ${\bf 2}_0+{\bf 1}_2$  & $A_2+E$      & Yes \\
  ${\bf 3}'''$   & $F_2$     & ${\bf 2}_0+{\bf 1}_3$  & $B_2+E$      &\\
  ${\bf 3}''''$  & $A_1+E$   & ${\bf 2}_4+{\bf 1}_4$  & $2A_1+B_1$   & Yes \\
  ${\bf 3}'''''$ & $A_2+E$   & ${\bf 2}_4+{\bf 1}_5$  & $A_1+2B_1$   &\\
  ${\bf 6}$      & $F_1+F_2$ & ${\bf 2}_1+{\bf 2}_3+{\bf 2}_5$ 
                                                      & $A_2+B_2+2E$ & Yes \\
  \hline
  \end{tabular}\end{center}
  \caption[dummy]{Irreps of $G$, defined in eq.\ (\ref{mes.grf}), and their
     reduction at finite temperature to irreps of $\check G$, defined in
     eq.\ (\ref{mes.grft}). The irreps which are realised for mesons are
     marked. Meson states do not exhaust all the irreps of $G$, $O$ or
     $\check G$, but do exhaust all the irreps of $D_4$.}
\label{tb.meson}\end{table}

In a meson representation, the quark number $q=0$. As a result, the
representants $X_k$ of the flavour generators $\Xi_k$ commute.
Consequently $C$ and $I$ commute with $X_k$. At $T=0$, it has been
shown that \cite{golt2}
\begin{equation}
   RF\;=\;GRF\left(X_\mu,R_{kl},I\right)\otimes Z_2(C),
\label{mes.first}\end{equation}
where
\begin{equation}
  GRF\left(X_\mu,R_{kl},I\right)\;=\;
       G\left(\tilde X_k,R_{kl}\right)\otimes Z_2(I)
            \otimes Z_2(X_1X_2X_3)\otimes Z_2(X_4),
\label{mes.grf}\end{equation}
with $\tilde X_k=X_kX_1X_2X_3$. The irreps of $GRF$ are denoted
$\r^{\sigma_4\sigma_{123}}$, where $\r$ denotes an irrep of $G$, and
$\sigma_4$ and $\sigma_{123}$ are signs which denote the irreps of the
$Z_2$ factor groups generated by $X_4$ and $X_1X_2X_3$ respectively.

Next we identify the group $G$. The $\tilde X_k$ generate a 4 element
Abelian group called the Viergruppe, $V=Z_2\otimes Z_2$. This is a normal
subgroup of $G$. The transformation properties of $\tilde X_k$ under
rotations, eq.\ (\ref{rev.flav}), show that G is the semi-direct product
$G=V(\tilde X_k)\semidir O$. Now, the cubic group $O=V(R_{kl}^2)\semidir S_3$,
where the normal subgroup, $V(R_{kl}^2)$ is generated by the three
rotations by angle $\pi$ \cite{simon}, and the other
factor is the permutation group of 3 elements. From eq.\ (\ref{rev.flav})
it is clear that $V(R_{kl}^2)$ has trivial action on $V(\tilde X_k)$, and
we can write
\begin{equation}
   G(\tilde X_k,R_{kl})\;=\;
        \biggl(V(\tilde X_k)\otimes V(R_{kl}^2)\biggr)\semidir S_3.
\label{mes.grfp}\end{equation}
Since the normal subgroup is Abelian, the irreps of $G$ can be efficiently
generated by the method of induced representations. Details are given
in Appendix \ref{ap.ind}, where we recover the results of \cite{golt2}.

\begin{table}[htbp]\begin{center}
  \begin{tabular}{|c|c|c|c|c|}  \hline
  $GRF$ & $\check G$ & $D_4^h$ & Operator \\
  \hline
  ${\bf 1}^{++}$ & ${\bf 1}_0$ & $A_1^+$ & $\sum_x\bar\chi(x)\chi(x)$ \\
  ${\bf 1}^{+-}$ & ${\bf 1}_0$ & $A_1^+$ & $\sum_x\eta_4(x)\zeta_4(x)\bar\chi(x)\chi(x)$ \\
  ${{\bf 3}''''}^{+-}$ & ${\bf 1}_4$ & $A_1^+$ &
      $\sum_x\epsilon(x)\eta_3(x)\zeta_3(x)\bar\chi(x)\chi(x)$ \\
                       & ${\bf 2}_4$ & $A_1^+$ &
      $\sum_x\epsilon(x)[\eta_1(x)\zeta_1(x)+\eta_2(x)\zeta_2(x)]\bar\chi(x)\chi(x)$ \\
                       & & $B_1^+$ &
      $\sum_x\epsilon(x)[\eta_1(x)\zeta_1(x)-\eta_2(x)\zeta_2(x)]\bar\chi(x)\chi(x)$ \\
  ${{\bf 3}''''}^{++}$ & ${\bf 1}_4$ & $A_1^+$ &
      $\sum_x\epsilon(x)\eta_4(x)\zeta_4(x)\eta_3(x)\zeta_3(x)\bar\chi(x)\chi(x)$ \\
                       & ${\bf 2}_4$ & $A_1^+$ &
      $\sum_x\epsilon(x)\eta_4(x)\zeta_4(x)[\eta_1(x)\zeta_1(x)+\eta_2(x)\zeta_2(x)]\bar\chi(x)\chi(x)$ \\
                       & & $B_1^+$ &
      $\sum_x\epsilon(x)\eta_4(x)\zeta_4(x)[\eta_1(x)\zeta_1(x)-\eta_2(x)\zeta_2(x)]\bar\chi(x)\chi(x)$ \\
  \hline
  \end{tabular}\end{center}
  \caption[dummy]{Representations of local staggered mesons. Only the $A_1^+$
     operators have been used in simulations till now. Reduction of three-link
     separated mesons follows an identical pattern and generates the opposite
     parity irreps of $D_4^h$.}
\label{tb.meson0}\end{table}

\begin{table}[htbp]\begin{center}
  \begin{tabular}{|c|c|c|c|c|}  \hline
  $GRF$ & $\check G$ & $D_4^h$ & Operator \\
  \hline
  ${\bf 3}^{-+}$ & ${\bf 1}_6$ & $A_2^-$ & $\sum_x\eta_3(x)\bar\chi(x)D_3\chi(x)$ \\
                 & ${\bf 2}_2$ & $E^-$   & $\sum_x\eta_{1,2}(x)\bar\chi(x)D_{1,2}\chi(x)$ \\
  ${\bf 3}^{--}$ & ${\bf 1}_6$ & $A_2^-$ &
      $\sum_x\eta_4(x)\zeta_4(x)\eta_3(x)\bar\chi(x)D_3\chi(x)$ \\
                 & ${\bf 2}_2$ & $E^-$   &
      $\sum_x\eta_4(x)\zeta_4(x)\eta_{1,2}(x)\bar\chi(x)D_{1,2}\chi(x)$ \\
  ${{\bf 3}''}^{--}$ & ${\bf 1}_2$ & $A_2^-$ &
      $\sum_x\epsilon(x)\zeta_3(x)\bar\chi(x)D_3\chi(x)$ \\
                     & ${\bf 2}_0$ & $E^-$ &
      $\sum_x\epsilon(x)\zeta_{1,2}(x)\bar\chi(x)D_{1,2}\chi(x)$ \\
  ${{\bf 3}''}^{-+}$ & ${\bf 1}_2$ & $A_2^-$ &
      $\sum_x\eta_4(x)\zeta_4(x)\epsilon(x)\zeta_3(x)\bar\chi(x)D_3\chi(x)$ \\
                     & ${\bf 2}_0$ & $E^-$ &
      $\sum_x\eta_4(x)\zeta_4(x)\epsilon(x)\zeta_{1,2}(x)\bar\chi(x)D_{1,2}\chi(x)$ \\
  ${\bf 6}^{--}$ & ${\bf 2}_5$ & $A_2^-$ &
      $\sum_x\epsilon(x)[\eta_1(x)+\eta_2(x)]\eta_3(x)\bar\chi(x)D_3\chi(x)$ \\
                 & & $B_2^-$ &
      $\sum_x\epsilon(x)[\eta_1(x)-\eta_2(x)]\eta_3(x)\bar\chi(x)D_3\chi(x)$ \\
                 & ${\bf 2}_1$ & $E^-$ &
      $\sum_x\epsilon(x)\eta_1(x)\eta_2(x)\bar\chi(x)D_2\chi(x)$,
      $1\leftrightarrow2$ \\
                 & ${\bf 2}_3$ & $E^-$ &
      $\sum_x\epsilon(x)\eta_3(x)\eta_1(x)\bar\chi(x)D_1\chi(x)$,
      $1\to2$ \\
  ${\bf 6}^{-+}$ & ${\bf 2}_5$ & $A_2^-$ &
      $\sum_x\epsilon(x)\eta_4(x)\zeta_4(x)[\eta_1(x)+\eta_2(x)]\eta_3(x)\bar\chi(x)D_3\chi(x)$ \\
                 & & $B_2^-$ &
      $\sum_x\epsilon(x)\eta_4(x)\zeta_4(x)[\eta_1(x)-\eta_2(x)]\eta_3(x)\bar\chi(x)D_3\chi(x)$ \\
                 & ${\bf 2}_1$ & $E^-$ &
      $\sum_x\epsilon(x)\eta_4(x)\zeta_4(x)\eta_1(x)\eta_2(x)\bar\chi(x)D_2\chi(x)$,
      $1\leftrightarrow2$ \\
                 & ${\bf 2}_3$ & $E^-$ &
      $\sum_x\epsilon(x)\eta_4(x)\zeta_4(x)\eta_3(x)\eta_1(x)\bar\chi(x)D_1\chi(x)$,
      $1\to2$ \\
  \hline
  \end{tabular}\end{center}
  \caption[dummy]{Representations of one-link separated staggered mesons. Here
     $D_\mu\phi({\bf x})=\phi({\bf x}+\hat\mu)+\phi({\bf x}-\hat\mu)$. Reduction
     of two-link separated mesons follows an identical pattern and generates
     the opposite parity irreps of $D_4^h$.}
\label{tb.meson1}\end{table}

This method makes it easy to construct the $T>0$ group,
\begin{equation}
  \underline{GRF}\;=\;\check G\otimes Z_2(I)
            \otimes Z_2(X_1X_2X_3)\otimes Z_2(X_4).
\label{mes.grft}\end{equation}
The rotation generators in $\check G$ are $R_{12}$ and $R_{13}^2$,
and they generate the group $D_4$. Since $D_4=V(R_{kl}^2)\semidir Z_2(R_{12})$,
we have
\begin{equation}
   \check G\;=\;
    \biggl(V(\tilde X_k)\otimes V(R_{kl}^2)\biggr)\semidir Z_2(R_{12})
\label{mes.grftp}\end{equation}
$\check G$ has 32 elements in 14 conjugacy classes. The irreps can
be constructed by the method of induced representations (see Appendix
\ref{ap.ind}). There are 8 one-dimensional and 6 two-dimensional irreps
of $\check G$.

The content of the various GRF irreps is shown in Table \ref{tb.meson}.
The reduction of irreps of $G$ to those of $O$, $\check G$ and $D_4$ are
performed using the character tables in Appendix \ref{ap.char}. These
reductions are consistent with those given in eq.\ (\ref{rev.oh2d4h}).
The irreps obtained for mesonic (quark-antiquark) operators can be
identified through the Clebsch-Gordan series for the GRF of Fermionic
representations.

We demonstrate the reduction of the ${\bf 1}^{+\pm}$ and
${{\bf 3}''''}^{+\pm}$ irreps of GRF to the irreps of $D_4^h$ in Table
\ref{tb.meson0}, using the full set of local meson operators. All the
operators which have been used to date for computing screening masses
belong to the $D_4^h$ irrep $A_1^+$, and conversely, all the $A_1^+$
operators in Table \ref{tb.meson0} have been used in measurements.
Notice, however, that this one irrep of $\underline{TS}$ descends from
different irreps of $\underline{GRF}$. The two $A_1^+$ irreps descending
from ${\bf 1}_0$ must give degenerate masses\footnote{This is the
phenomenon of ``parity doubling'' at high temperature.}, as must the
pairs descending from the ${\bf 1}_4$ and the ${\bf 2}_4$. However, there
is no group theoretical necessity for the three pairs to have the same
mass.

The reduction of ${\bf 3}^{-\pm}$, ${{\bf 3}''}^{-\pm}$ and ${\bf 6}^{-\pm}$
irreps obtained for one-link separated meson operators is given in Table
\ref{tb.meson1}. Reductions of two-link separated meson operators can also
be read off from the structure of these reductions. The latter give
positive $I$ parity irreps of $D_4^h$. Combining these two sets we have
a set of $A_2^\pm$, $B_2^\pm$ and $E^\pm$ irreps. Three link separated
operators reduce in the same way as the local meson operators but give
the opposite $I$ parity. These two sets together give us the remaining
irreps of $D_4^h$, {\sl i.e.\/}, $A_1^\pm$ and $B_1^\pm$.

\subsection{\label{sc.qk}Quark and Baryon Operators}

\begin{table}[htbp]\begin{center}
  \begin{tabular}{|c|c|c|c|}  \hline
  $GRF$ & $O_h$ & $D_4^h$ & Component \\
  \hline
  ${\bf 8}$ & $A_1^+$ & $A_1^+$ & $\chi(0)$ \\
   & $A_1^-$ & $A_1^-$ & $\chi(\hat x+\hat y+\hat z)$ \\
   & $F_1^+$ & $A_2^+$ & $\chi(\hat x+\hat y)$ \\
   &    & $E^+$ & $\left\{\chi(\hat x+\hat z),-\chi(\hat y+\hat z)\right\}$ \\
   & $F_1^-$ & $A_2^-$ & $\chi(\hat z)$ \\
   &    & $E^-$ & $\left\{\chi(\hat x),\chi(\hat y)\right\}$ \\
  \hline
  \end{tabular}\end{center}
  \caption[dummy]{Representations of the quark field $\chi_A$. The 
     ``zero-momentum'' projection is performed as shown in eq.\ 
     (\ref{rev.field}). The ${\bf 8}$ of $G_F$ is the same as the
     ${\bf 8}_0$ of the ${\underline G}_F$}
\label{tb.quark}\end{table}

Quark and baryon operators carry odd Fermion charge. The representants
of $\Xi_\mu$ anti-commute and generate the 32 element Clifford group
$CL(4)$. Its commutator subgroup is isomorphic to $Z_2$ and its Abelisation,
$CL(4)/Z_2$ is precisely the group $V(\tilde X_k)\otimes Z_2(X_4)\otimes
Z_2(X_1X_2X_3)$ encountered as the normal subgroup of the mesonic
rotation-shift group \cite{simon}. Apart from
the 16 one-dimensional irreps of this group, $CL(4)$ also has a
four-dimensional quaternionic irrep familiar to us from the algebra
of the Dirac matrices.

The rotation-shift group for Fermionic operators at $T=0$ is
\begin{equation}
   G_F\;=\;CL(4)\semidir O\;=\;
      \left(CL(4)\otimes V(R_{kl}^2)\right)\semidir S_3.
\label{qk.grf}\end{equation}
This group has 45 conjugacy classes, and hence 45 irreps. 40 of these
have been identified as the irreps $\r^{\sigma_4\sigma_{123}}$ of the
mesonic rotation-shift group. The remaining 5 are obtained by
inducing with the remaining non-trivial irrep of $CL(4)$. This gives
the five real irreps $\bf 8$, ${\bf 8}'$, $\bf 16$, $\bf 24$ and
${\bf 24}'$.

The defining representation of $G_F$, $\bf 8$, is given by zero momentum
staggered Fermions on a time slice eq.\ (\ref{rev.zerom}). Under $O_h$ the
octet breaks \cite{golt1}
\begin{equation}
   {\bf 8}\;\to\; A_1^+ + A_1^- + F_1^+ + F_1^-.
\label{qk.break0}\end{equation}
The Clebsch-Gordans for ${\bf 8}\times{\bf 8}\times{\bf 8}$ show that
only the $\bf 8$, $\bf 8'$ and $\bf 16$ are found as irreps of baryons
\cite{golt1}. The local baryon operators built from staggered Fermions
transform as the $A_1^+$ component of the $\bf 8$.

The rotation-shift group for Fermionic operators at $T>0$ is
\begin{equation}
   {\underline G}_F\;=\;CL(4)\semidir D_4\;=\;
      \left(CL(4)\otimes V(R_{kl}^2)\right)\semidir Z_2.
\label{qk.grft}\end{equation}
This group has 61 conjugacy classes, and hence 61 irreps. 56 of these
have been identified as the irreps $\r^{\sigma_4\sigma_{123}}$ of the
$T>0$ mesonic rotation-shift group. The remaining 5 are obtained by
inducing with the remaining non-trivial irrep of $CL(4)$. This gives
the five real irreps ${\bf 8}_0$, ${\bf 8}_1$, ${\bf 8}_2$, ${\bf 8}_3$,
$\bf 16$. The reductions of the $T=0$ baryon irreps at $T>0$ are---
\begin{equation}
   {\bf 8}\to{\bf 8}_0,\qquad
   {\bf 8}'\to{\bf 8}_2,\qquad
   {\bf 16}\to{\bf 8}_0+{\bf 8}_2.
\label{qk.reduc}\end{equation}
Under $D_4^h$, we find that
\begin{equation}
   {\bf 8}_0\;\to\; A_1^+ + A_1^- + A_2^+ + A_2^- + E^+ + E^-.
\label{qk.break1}\end{equation}
The components of the quark field which carry different representations of
$TS$ and $\underline{TS}$ are shown in Table \ref{tb.quark}.

Given the classification of baryon operators in \cite{golt1} it is a
simple matter to construct the $D_4^h$ irreps from them. We do not present
a detailed table, because the state of the art in measurements with baryons
has not progressed far beyond the measurement of the purely local
operators even at $T=0$.

\section{\label{sc.fft}Free Field Theory and Beyond}

In a free field theory of staggered Fermions, the symmetries of the Hamiltonian
become symmetries of the field configuration. As a result, there are many more
degeneracies among the screening masses than in the general problem. The only
relevant group turns out to be $\underline{TS}$, in the sense that all
correlators in the same irrep of $\underline{TS}$ have degenerate masses, even
if they descend through different irreps of $\underline{RF}$ and
$\underline{GRF}$. However, the degeneracies are even higher than would be
predicted by the application of the group theory of $\underline{TS}$.

The local ``mesons'' in the $A_1^+$ irrep of $D_4^h$ have been analysed
extensively in free field theory (FFT) \cite{fss,myold}. It is known that
the correlation functions show the typical even-odd structure of staggered
Fermion correlators. The screening mass, $\mu$, in this channel is
\begin{equation}
   \mu a = 2\sqrt{m^2 a^2 + \sin^2\left(\frac{\pi}{N_\tau}\right)}
       \to \frac{2\pi}{N_\tau} = 2\pi T a.
\label{fft.mass}\end{equation}
Here $a$ is the lattice spacing, $N_\tau$ is the lattice size in the
Euclidean time direction, and $m$ is the quark mass. The limit is taken
for small $m$ and large $N_\tau$, and is equal to twice the
minimum Matsubara frequency, $2\pi T$ \cite{screen,fss,elio}. Finite
size effects are clearly strong, even in FFT, and have been analysed
before \cite{fss,myold}. 

The remaining local ``mesons'' are in the $B_1^+$ irrep of $D_4^h$. In
FFT these correlation functions vanish. This is easy to understand. In FFT
the $x$ and $y$ direction propagators are exactly equivalent, and hence
their difference (see Table \ref{tb.meson0}) cancels.

The non-local mesons also divide into two groups. The $A_1^\pm$, $A_2^\pm$
and $E^\pm$ irreps of $D_4^h$ give rise to screening masses according to
eq.\ (\ref{fft.mass}). In contrast, the $B_1^\pm$ and $B_2^\pm$ irreps vanish
in FFT. Deviations from any of these free field theory results can be used
as a measure of the interaction strength between Fermions.

It is interesting to recall past measurements \cite{screen,fss,myold}.
Screening masses have been measured with four correlators---
\begin{itemize}
\item 
   the $D_4^h$ irrep $A_1^+$ descending from the ${\bf 1}_0^{+\pm}$ 
   of the $\underline{GRF}$, and
\item
   a linear combination of the two pairs of $D_4^h$ irreps $A_1^+$ descending
   from the ${\bf 1}_4^{+\pm}$ and ${\bf 2}_4^{+\pm}$ of the $\underline{GRF}$
   (see Table \ref{tb.meson0}).
\end{itemize}
Since they all belong to the same irrep of $\underline{TS}$, we might expect
them to give the same screening mass in FFT. Extensive numerical work was
performed in the 4-flavour \cite{fss} and quenched \cite{myold} $SU(3)$
theories at $T\approx T_c$, $T=3T_c/2$ and $T=2T_c$. It was found that the
masses within each group become degenerate already quite close to $T_c$, but
the masses of the two groups differed by about 10\% even at $2T_c$.
In the light of the preceding calculations, the natural explanation for this
observation is that there are residual interactions.

A minor controversy persists in the interpretation of these observations.
The two viewpoints can be summarised as---
\begin{enumerate}
\item
   Fermions in QCD at $T>T_c$ are weakly coupled, since the $A_1^+$ screening
   masses coming from the ${\bf 1}_4^{+\pm}$ and the ${\bf 2}_4^{+\pm}$ agree
   very well with eq.\ (\ref{fft.mass}). This picture becomes a better
   approximation with increasing $T$.
\item
   QCD at $T_c<T<2T_c$ is not very different from that at $T<T_c$, since the
   $A_1^+$ from the ${\bf 1}_0^{+\pm}$ is quite different from
   eq.\ (\ref{fft.mass}). Furthermore, the $A_1^+$ from the ${\bf 1}_4^{+\pm}$
   and the ${\bf 2}_4^{+\pm}$ are just the same as the vector/pseudo-vector
   at zero temperature.
\end{enumerate}
To resolve this controversy one needs first to measure the $B_1^+$ coming from
the ${\bf 2}_4^{+\pm}$ and check whether it agrees with either of the models
above. Beyond this, one needs to measure screening masses in other irreps of
the $\underline{GRF}$ which are in the same irreps of the continuum group as
the measured local Fermion bilinears. This would check whether the second model
is viable or not.

\section{\label{sc.end}Summary}

In this paper we have studied the symmetries of the transfer matrix for
staggered Fermions. By using the method of induced representations we have
reproduced old, known results for the spin-flavour symmetry group and its
irreps at zero temperature \cite{golt1,golt2}. This method allows a simple
generalisation to the analysis of the spatial direction transfer matrix at
finite temperature. Our main result is the identification of the irreps of
the finite temperature symmetry group given in Table \ref{tb.char2}.

Using this, we have decomposed the $T=0$ irreps into the $T>0$ irreps. This
reduction is shown for ``mesons'' in Tables \ref{tb.meson0} and
\ref{tb.meson1}. For quarks the reduction is given in Table \ref{tb.quark}.
The reduction for local ``baryons'' may be read off from the same table.

The phenomenon of ``parity doubling'' at high temperatures is allowed by the
group theory. In a free fermion theory, in fact, the degeneracies are much
higher--- the screening masses are classified by the point group of the
spatial slice of the lattice. In an interacting theory this is not true;
the descent of each irrep through the chain of enveloping groups (see
eq.\ \ref{rev.tchain}) is important. This has been seen in the $A_1^+$ irrep
of the point group. Whether the physics at $T>T_c$ is a small perturbation
around the free theory or the zero temperature theory (or something else
altogether) can be explored by studying other irreps. The $B_1^+$ is a good
candidate because it is also built from local fermion bilinear operators.

I would like to thank F.\ Karsch and E.\ Laermann for discussions when this
work was started, and T.\ Venkataramanna for patient tutorials on the use of
the theory of induced representations.

\newpage\appendix
\section{\label{ap.ind}Induced Representations}

\begin{table}[htbp]\begin{center}
  \begin{tabular}{|c|c|c|c|}  \hline
  Orbit & Isotropy & Dimension & Multiplicity \\
  \hline
  $\underline{(0,0)}$ & $S_3$ & 1 & 2 \\
          &     & 2 & 1 \\
  $\underline{(0,1)},(0,2),(0,3)$ & $Z_2(R_{12})$ & 3 & 2 \\
  $\underline{(1,0)},(2,0),(3,0)$ & $Z_2(R_{12})$ & 3 & 2 \\
  $\underline{(1,1)},(2,2),(3,3)$ & $Z_2(R_{12})$ & 3 & 2 \\
  $\underline{(1,2)},(1,3),(2,1),$ & & & \\
               $(2,3),(3,1),(3,2)$ & $\{E\}$ & 6 & 1 \\
  \hline
  \end{tabular}\end{center}
  \caption[dummy]{Induced representations of $G=\left(v\otimes V\right)
      \semidir S_3$. A representative member of each orbit is underlined.
      The trivial isotropy group is denoted by $\{E\}$. This construction
      reproduces the result of \cite{golt2}.}
\label{tb.inds3}\end{table}

The method of induced representations for a semi-direct product group
$G=N\semidir H$, for Abelian $N$, can be found in \cite{simon}. Here
we quote the results required in this paper. The dual, $\hat N$ (set
of equivalence classes of irreps of $N$), is isomorphic to $N$, and
the action of $H$ on $\hat N$ is isomorphic to its action on $N$.

\begin{table}[htbp]\begin{center}
  \begin{tabular}{|c|c|c|c|}  \hline
  Orbit & Isotropy & Dimension & Multiplicity \\
  \hline
  $\underline{(0,0)}$ & $Z_2$ & 1 & 2 \\
  $\underline{(0,3)}$ & $Z_2$ & 1 & 2 \\
  $\underline{(3,0)}$ & $Z_2$ & 1 & 2 \\
  $\underline{(3,3)}$ & $Z_2$ & 1 & 2 \\
  $\underline{(0,1)},(0,2)$ & $\{E\}$ & 2 & 1 \\
  $\underline{(1,0)},(2,0)$ & $\{E\}$ & 2 & 1 \\
  $\underline{(1,1)},(2,2)$ & $\{E\}$ & 2 & 1 \\
  $\underline{(1,2)},(2,1)$ & $\{E\}$ & 2 & 1 \\
  $\underline{(1,3)},(2,3)$ & $\{E\}$ & 2 & 1 \\
  $\underline{(3,1)},(3,2)$ & $\{E\}$ & 2 & 1 \\
  \hline
  \end{tabular}\end{center}
  \caption[dummy]{Induced representations of $\check G=\left(V\otimes V\right)
      \semidir Z_2$. A representative member of each orbit is underlined.
      The trivial isotropy group is denoted by $\{E\}$.}
\label{tb.inds2}\end{table}

Under the action of $H$, the dual breaks up into disjoint orbits $O_i$,
{\sl i.e.\/}, $\hat N=\oplus_i O_i$. Examine the isotropy group,
$H_i\subset H$ of one representative $\chi_i\in O_i$. We need to know
two cases---
\begin{itemize}
\item
   When the orbit $O_i$ has only one element, {\sl i.e.\/}, $H_i=H$,
   then the induced representations are precisely the irreps of $H$.
\item
   For other $\chi_i$ when $H_i$ is Abelian, the dimension of the
   induced representation is the number of elements in the coset
   $H/H_i$, and the multiplicity of such irreps is given by the
   number of classes in $H_i$.
\end{itemize}

All 4 irreps of the Viergruppe, $V$, are one-dimensional. We label them
by the numbers 0, 1, 2 and 3. The trivial irrep is called 0. The irrep
labelled $k$ ($\ne0$) has $\chi(E)=\chi(X_k)=1$ and the other two
characters $-1$. Irreps
of the direct product $V\otimes V$ are labelled by the ordered pair
$(k,l)$ where $k$ is an irreps of the first factor and $l$ of the second.
$S_3$ has two one-dimensional irreps (the trivial and the sign) and one
two-dimensional irrep. $Z_2$ has two one-dimensional irreps, the trivial
and the sign. These are the only inputs into the construction of the
irreps we require. The construction of the irreps of $G$ and $\check G$
(see eqs.\ \ref{mes.grfp} and \ref{mes.grftp}), follow from the rules above,
and are given in Tables \ref{tb.inds3} and \ref{tb.inds2} respectively.

\section{\label{ap.char}Character Tables}

\begin{table}[htbp]\begin{center}
  \begin{tabular}{|c|c|r|r|r|r|r|r|r|r|r|r|}  \hline
  $G$ & $O$ & $E$ & $u$ & $v$ & $u\cdot v$ &
              $uv$ & $\cal C$ & $R$ &
              $uR$ & $vR$ & $uvR$ \\
  \cline{3-12} 
  & & 1 & 3 & 3 & 3 & 6 & 32 & 12 & 12 & 12 & 12 \\
  \hline
  ${\bf 1}$ & $A_1$ &
        $1$ & $1$ & $1$ & $1$ & $1$ & $1$ & $1$ & $1$ & $1$ & $1$ \\
  ${\bf 1}'$ & $A_2$ &
        $1$ & $1$ & $1$ & $1$ & $1$ & $1$ & $-1$ & $-1$ & $-1$ & $-1$ \\
  ${\bf 2}$ & $E$ &
        $2$ & $2$ & $2$ & $2$ & $2$ & $-1$ & $0$ & $0$ & $0$ & $0$ \\
  ${\bf 3}$ & $F_1$ &
        $3$ & $3$ & $-1$ & $-1$ & $-1$ & $0$ & $1$ & $1$ & $-1$ & $-1$ \\
  ${\bf 3}'$ & $F_2$ &
        $3$ & $3$ & $-1$ & $-1$ & $-1$ & $0$ & $-1$ & $-1$ & $1$ & $1$ \\
  ${\bf 3}''$ & $F_1$ &
        $3$ & $-1$ & $-1$ & $3$ & $-1$ & $0$ & $1$ & $-1$ & $-1$ & $1$ \\
  ${\bf 3}'''$ & $F_1$ &
        $3$ & $-1$ & $-1$ & $3$ & $-1$ & $0$ & $-1$ & $1$ & $1$ & $-1$ \\
  ${\bf 3}''''$ & $E+A_1$ &
        $3$ & $-1$ & $3$ & $-1$ & $-1$ & $0$ & $1$ & $-1$ & $1$ & $-1$ \\
  ${\bf 3}''''$ & $E+A_2$ &
        $3$ & $-1$ & $3$ & $-1$ & $-1$ & $0$ & $-1$ & $1$ & $-1$ & $1$ \\
  ${\bf 6}$ & $F_1+F_2$ &
        $6$ & $-2$ & $-2$ & $-2$ & $2$ & $0$ & $0$ & $0$ & $0$ & $0$ \\
  \hline
  \end{tabular}\end{center}
  \caption[dummy]{The character table for $G=\left(V(u)\otimes V(v)\right)
      \semidir S_3$. The second line of the table gives the number of
      operators in each class.}
\label{tb.char1}\end{table}

\begin{table}[htbp]\begin{center}
  \begin{tabular}{|c|c|r|r|r|r|r|r|r|r|r|r|r|r|r|r|}  \hline
  $\check G$ & $D_4$ & $E$ & $u$ & $v$ & $uv$ & $\bf u$ & $\bf v$ &
                       $u\bf v$ & $v\bf u$ & $\bf u\cdot v$ & $\bf uv$ &
                       $R$ & ${\bf u}R$ & ${\bf v}R$ & ${\bf uv}R$ \\
  \cline{3-16} 
  & & 1 & 1 & 1 & 1 & 2 & 2 & 2 & 2 & 2 & 2 & 4 & 4 & 4 & 4 \\
  \hline
  ${\bf 1}_0$ & $A_1$ &
    $1$ & $1$ & $1$ & $1$ & $1$ & $1$ & $1$ & $1$ & $1$ & $1$ &
    $1$ & $1$ & $1$ & $1$ \\
  ${\bf 1}_1$ & $B_1$ &
    $1$ & $1$ & $1$ & $1$ & $1$ & $1$ & $1$ & $1$ & $1$ & $1$ &
    $-1$ & $-1$ & $-1$ & $-1$ \\
  ${\bf 1}_2$ & $A_2$ &
    $1$ & $1$ & $1$ & $1$ & $-1$ & $-1$ & $-1$ & $-1$ & $1$ & $1$ &
    $1$ & $-1$ & $-1$ & $1$ \\
  ${\bf 1}_3$ & $B_2$ &
    $1$ & $1$ & $1$ & $1$ & $-1$ & $-1$ & $-1$ & $-1$ & $1$ & $1$ &
    $-1$ & $1$ & $1$ & $-1$ \\
  ${\bf 1}_4$ & $A_1$ &
    $1$ & $1$ & $1$ & $1$ & $-1$ & $1$ & $1$ & $-1$ & $-1$ & $-1$ &
    $1$ & $-1$ & $1$ & $-1$ \\
  ${\bf 1}_5$ & $B_1$ &
    $1$ & $1$ & $1$ & $1$ & $-1$ & $1$ & $1$ & $-1$ & $-1$ & $-1$ &
    $-1$ & $1$ & $-1$ & $1$ \\
  ${\bf 1}_6$ & $A_2$ &
    $1$ & $1$ & $1$ & $1$ & $1$ & $-1$ & $-1$ & $1$ & $-1$ & $-1$ &
    $1$ & $1$ & $-1$ & $-1$ \\
  ${\bf 1}_7$ & $B_2$ &
    $1$ & $1$ & $1$ & $1$ & $1$ & $-1$ & $-1$ & $1$ & $-1$ & $-1$ &
    $-1$ & $-1$ & $1$ & $1$ \\
  ${\bf 2}_0$ & $E$ &
    $2$ & $-2$ & $-2$ & $2$ & $0$ & $0$ & $0$ & $0$ & $2$ & $-2$ &
    $0$ & $0$ & $0$ & $0$ \\
  ${\bf 2}_1$ & $E$ &
    $2$ & $-2$ & $-2$ & $2$ & $0$ & $0$ & $0$ & $0$ & $-2$ & $2$ &
    $0$ & $0$ & $0$ & $0$ \\
  ${\bf 2}_2$ & $E$ &
    $2$ & $2$ & $-2$ & $-2$ & $2$ & $0$ & $0$ & $-2$ & $0$ & $0$ &
    $0$ & $0$ & $0$ & $0$ \\
  ${\bf 2}_3$ & $E$ &
    $2$ & $2$ & $-2$ & $-2$ & $-2$ & $0$ & $0$ & $2$ & $0$ & $0$ &
    $0$ & $0$ & $0$ & $0$ \\
  ${\bf 2}_4$ & $A_1+B_1$ &
    $2$ & $-2$ & $2$ & $-2$ & $0$ & $2$ & $-2$ & $0$ & $0$ & $0$ &
    $0$ & $0$ & $0$ & $0$ \\
  ${\bf 2}_5$ & $A_2+B_2$ &
    $2$ & $-2$ & $2$ & $-2$ & $0$ & $-2$ & $2$ & $0$ & $0$ & $0$ &
    $0$ & $0$ & $0$ & $0$ \\
  \hline
  \end{tabular}\end{center}
  \caption[dummy]{The character table for $\check G=\left(V(u)\otimes V(v)
      \right)\semidir Z_2$. The second line of the table gives the number
      of operators in each class.}
\label{tb.char2}\end{table}

In this appendix we define the irreps of $G$ and $\check G$ (see eqs.\ 
\ref{mes.grfp}, \ref{mes.grftp}) by writing down the character tables.

In Appendix \ref{ap.ind} we showed that the group $G$ of eq.\ (\ref{mes.grfp})
has 10 irreps. Hence the 96 elements of the group fall into 10 conjugacy
classes. For every $g\in G$, we can write uniquely $g=uvs$, where
$u\in V(\tilde X_k)$, $v\in V(R_{kl}^2)$, and $s\in S_3$. We use the notation
$R_{kl}$ and ${\cal C}$ respectively for the operators in $S_3$ which permute
the pair $kl$ and make a cyclic shift. In terms of this decomposition, the 10
conjugacy classes are--- the identity E, $u_i$ (with $i=1,2,3$), $v_i$,
$u\cdot v$ (meaning by $u_iv_i$), $u_iv_j$ (with $j\ne i$, and denoted $uv$),
$\cal C$, $R_{kl}$, $u_iR_{ij}$ (denoted $uR$), $v_iR_{ij}$ (denoted $vR$),
$u_iv_jR_{ij}$ (denoted $uvR$). The character table is constructed by standard
methods, and given in Table \ref{tb.char1}.
The reduction of irreps of $G$ to that of the subgroup of cubic rotations
$O$ is performed by inspecting the characters\footnote{Character tables for
$O_h$ and $D_4^h$ may be found in \cite{hammer}}  of the conjugacy classes of
$O$. These are the rotations by $\pi/2$ ($R$), rotations by $\pi$ ($v$),
rotations by $2\pi/3$ ($\cal C$) and the remaining 2-fold rotations
($vR$).

The construction, in Appendix \ref{ap.ind}, of the irreps of the group
$\check G$ of eq.\ (\ref{mes.grftp}) shows that there are 14 conjugacy
classes. For every $g\in\check G$ we have the unique decomposition
$g=uvs$, where $u\in V(\tilde X_k)$, $v\in V(R_{kl}^2)$, and $s\in Z_2$.
Using the notation $R$ for the nontrivial element of $Z_2$, we can
write the conjugacy classes as $E$, $u=u_3$, $v=v_3$, $u_i$ (with $i=1,2$
and denoted $\bf u$), $\bf v$ (in the same notation), $uv$, $u_iv_i$
(with $i=1,2$ and denoted $\bf u\cdot v$), $u\bf v$, $v\bf u$,
$\bf uv$ (meaning $u_iv_j$ with $i\ne j=1,2$), $R$ (which is equivalent
to $uvR$, $uR$ and $vR$), ${\bf u}R$ (also equivalent to $v{\bf u}R$),
${\bf v}R$ (also equivalent to $u{\bf v}R$), and ${\bf uv}R$. The
character table is given in Table \ref{tb.char2}. The decomposition to
irreps of $D_4$ needs the identification $R_{12}^2=v$, the $\pi$
rotations about the $x$ and $y$ axes are $\bf v$, $R_{12}=R$ and the
remaining 2-fold symmetries are ${\bf v}R$.

\end{document}